\AtBeginDocument{%
  \mathchardef\mathcomma\mathcode`\,
  \mathcode`\,="8000 
}
{\catcode`,=\active
  \gdef,{\mathcomma\discretionary{}{}{}}
}
\documentclass[final,5p,times,twocolumn]{elsarticle}

\usepackage{amssymb}
\usepackage{amsmath}
\usepackage{amsthm}

\usepackage{xcolor}
\usepackage{color,soul}
\def\HiLi{\leavevmode\rlap{\hbox to \hsize{\color{yellow!}\leaders\hrule height .8\baselineskip depth .5ex\hfill}}}

\usepackage{listings}
\usepackage[ruled,vlined]{algorithm2e}
\SetEndCharOfAlgoLine{}

\journal{Information and Software Technology}

\begin{document}

\begin{frontmatter}


\title{Source Code Optimization using Equivalent Mutants\tnoteref{acks}}
\tnotetext[acks]{{\textcopyright} 2018. This manuscript version is made available under the CC-BY-NC-ND 4.0 license http://creativecommons.org/licenses/by-nc-nd/4.0/.}
\author[af1]{Jorge L\'opez}
\ead{jorge.lopez@telecom-sudparis.eu}

\author[af1]{Natalia Kushik}
\ead{natalia.kushik@telecom-sudparis.eu}

\author[af2]{Nina Yevtushenko}
\ead{evtushenko@ispras.ru}
\address[af1]{SAMOVAR, CNRS, T\'el\'ecom SudParis, Universit\'e Paris-Saclay, 9 rue Charles Fourier, 91000 \'Evry, France}

\address[af2]{Ivannikov Institute for System Programming of the Russian Academy of Sciences, 25 Alexander Solzhenitsyn street, 109004, Moscow, Russia}



\begin{abstract}\label{abstract}

\noindent\textbf{Context:} A \emph{mutant} is a program obtained by syntactically modifying a program's source code; an \emph{equivalent mutant} is a mutant, which is functionally equivalent to the original program. Mutants are primarily used in \emph{mutation testing}, and when deriving a test suite, obtaining an equivalent mutant is considered to be highly negative, although these equivalent mutants could be used for other purposes. 

\noindent\textbf{Objective:} We present an approach that considers equivalent mutants valuable, and utilizes them for \emph{source code optimization}. Source code optimization \emph{enhances} a program's source code preserving its behavior. 

\noindent\textbf{Method:} We showcase a procedure to achieve source code optimization based on equivalent mutants and discuss proper \emph{mutation operators}. 

\noindent\textbf{Results:} Experimental evaluation with Java and C programs demonstrates the applicability of the proposed approach. 

\noindent\textbf{Conclusion:} An algorithmic approach for source code optimization using equivalent mutants is proposed. It is showcased that whenever applicable, the approach can outperform traditional compiler optimizations.

\end{abstract}


\end{frontmatter}


\section{Introduction}\label{intro}
\emph{Source code optimization} is a process which \emph{enhances} a program's source code, in order to obtain a functionally \emph{equivalent} program, i.e., a program which computes the same solution for the same problem but, possesses better non-functional aspects. Traditionally, source code optimization techniques are implemented on compilers \cite{dragonbook}.

Program mutants are used in mutation testing \cite{mujava}, a software testing technique whose main idea is to modify the original source code to obtain a \emph{mutant} that should be later distinguished from the original program by a test case. The program modification is performed using a \emph{mutation operator}; a mutation operator performs changes to the original source code. When applying a mutation operator, an equivalent program called an \emph{equivalent mutant} can be obtained. Mutation testing attempts to detect and avoid equivalent mutants \cite{mutantimpact}. We note that detecting equivalent mutants using compiler optimizations is well established~\cite{mutdetcomp2}. However, to the best of our knowledge, the first publication where \emph{a novel use of equivalent mutants} is discussed, appeared recently~\cite{noveluseofmutants}; the authors show that equivalent mutants can be used for static anomaly detection, e.g., to detect if the mutated code possesses better readability, better execution time, etc. However, the authors do not study nor outline a procedure where mutation operators are used for source code optimization. 

Equivalent mutants can provide an optimized source code in terms of its (program/binary) execution time and other aspects. However, to effectively use the software mutation technique for source code optimization, several questions should be addressed: what are the mutation operators which can provide such optimizations? how to apply such mutation operators for optimization purposes? what is the benefit of the mutation-based source code optimization compared to traditional source code optimization? This paper is devoted to answer these questions; further, we perform preliminary experiments with a mutation software, $\mu$Java~\cite{mujava}, which showcase the applicability and effectiveness of the proposed approach (Section~\ref{experiments}).


\section{Equivalent Mutants for Source Code Optimization}\label{contrib}
Given a (computer) program $\mathcal{P}$, we denote $\mathcal{S}_{\mathcal{P}}$ its associated source code. $\mathcal{P}$ is obtained from $\mathcal{S}_{\mathcal{P}}$ through a proper compilation process, i.e., a function $C:\Sigma^{*} \mapsto \{0,1\}^{*}$ that maps a program's source code (a string over a particular programming language alphabet $\Sigma$) into a binary (or executable) code, i.e., $\mathcal{P} = C(\mathcal{S}_{\mathcal{P}})$. We denote the set of all possible inputs for $\mathcal{P}$ as $I$; correspondingly, $O$ is the set of all possible outputs of $\mathcal{P}$. An input sequence is denoted as $\alpha\in I^*$; correspondingly, an output sequence $\beta \in O^{*}$ is the program's output response to this sequence, denoted as $out(\mathcal{P}, \alpha)$\footnote{We assume that the program is deterministic and, therefore, such output is unique.}. We consider a program's running time under a given input sequence $\alpha$, in a common and predefined architecture, measured in milliseconds~($ms$) and denoted as $t(\mathcal{P},\alpha)$. Correspondingly, we denote the \emph{overall running time of a program} $\mathcal{P}$ with respect to a set of input sequences $M$ as $\tau=\sum_{\alpha \in M}t(\mathcal{P},\alpha)$.

A program $\mathcal{P}$ is \emph{$M$-equivalent} to $\mathcal{P}'$ (written $\mathcal{P} \overset{M}{\equiv} \mathcal{P}'$) if $\forall \alpha \in M\; out(\mathcal{P}, \alpha)=out(\mathcal{P}',\alpha)$. We focus on program $M$-equivalence due to the fact that in the general case, the problem of checking the equivalence of two arbitrary programs is undecidable. However, in some cases, equivalence with respect to a finite set of inputs implies complete functional equivalence when having a behavior model \cite{mbt1}. Furthermore, many programs are used only within a context, receiving only a subset of possible (defined) inputs, or the program is only developed for a subset of inputs. Likewise, it is well-known that regression tests (a finite subset of the program inputs) are becoming an industry standard, and they somehow guarantee that a new version (including an optimized one) behaves as required.

A source code optimization process is a function which receives a source code and produces a new (optimized) source code $\mathcal{O}: \Sigma^{*}\mapsto\Sigma^{*}$. The obtained source code compiles to a functionally equivalent program with respect to an input set $M$, i.e., $C(\mathcal{S}_{\mathcal{P}}) \overset{M}{\equiv} \mathcal{O}(C(\mathcal{S}_{\mathcal{P}}))$. As it is not possible to derive an algorithmic approach to compute the time complexity of a program, optimality is considered with respect to the overall running time of a program, i.e, $\sum_{\alpha\in M}t(C(\mathcal{O}(\mathcal{S}_{\mathcal{P}})),\alpha) <\sum_{\alpha\in M}t(C(\mathcal{S}_{\mathcal{P}}),\alpha)$. 

Arcaini et al.~\cite{noveluseofmutants} showcased that mutants can be \textit{better} than the original source code, including the case when the mutated source code has better time complexity than the original one. However, no discussion was performed on how equivalent mutants can be exploited. Therefore, the problem stated and solved in this paper is as follows: how can source code optimization be forced by the use of source code mutation? It is important to highlight that the use of source code mutants to enhance the source code's non-functional properties is limited in the literature; for a comprehensive survey on the subject the interested reader can refer to \cite{mutatnssurvey}.

We assume that there exist certain mutation operators which are more likely to provide source code optimization due to their nature. Operators as statement deletion can optimize the source code by performing a \textit{dead code elimination}, arithmetic operator replacement can optimize the source code by performing operators' \textit{strength reduction}, etc. \cite{dragonbook}. Nevertheless, compiler optimizations are likely to be more effective while performed on target by a compiler. Therefore, the question arises: are there any mutation operators that can produce source code optimizations which are different from the known compiler optimizations? Indeed, we collected the following set of mutation operators based on the method-level mutation operators of $\mu$Java~\cite{mujava}:

\begin{itemize}
    \item Relational Operator Replacement~(ROR): replaces relational operators with others, e.g., \lstinline{>=} with \lstinline{>}. In certain cases, avoiding to execute the code when the condition reaches equality can enhance the performance (as shown in \cite{noveluseofmutants}), for example, when searching for the maximum number within an array as shown in the following code snippet (hereafter $\Delta$ denotes the difference/replacement, i.e., the obtained mutant).
    \begin{lstlisting}[ mathescape=true,,language=java]
for (int i = 0; i < arr.length(); i++)
   if (arr[i] >= max) 
$\Delta$   if (arr[i] $>$ max)
      max = arr[i];
    \end{lstlisting}
    \item Shortcut Assignment Operator Replacement~(ASR): replaces shortcut assignment operators with other shortcut assignment operators, e.g., \lstinline{+=} with \lstinline{*=}. In certain cases, advancing faster in the progression can avoid the execution of loop cycles, for example, when working over the powers of a given number as shown in the following code snippet.
    \begin{lstlisting}[mathescape=true,,language=java]
   for(int i = 1; i <= N; i+=3)
$\Delta$   for(int i = 1; i <= N; i*=3)
      if(i > 0 && 1162261467 % i == 0) 
         //If-body
    \end{lstlisting}
    \item Arithmetic Operator Replacement~(AOR): replaces arithmetic operators with others, e.g., from \lstinline{+} to \lstinline{*}; similar to ASR, AOR can help advancing faster in the progressions.
\end{itemize}

We are interested in the set of mutation operators that perform different optimizations from traditional compiler optimizations, and can be applicable to different programming languages. Let $\mu=\{ROR,ASR,AOR\}$ be the set of mutation operators of interest. This set can be always extended by adding other mutation operators that can also perform compiler optimizations. We aim at limiting the mutation operators to be considered in order to avoid deriving mutants that do not optimize the source code. Indeed, executing all mutants against the set of inputs $M$ may take a very long time. However, we note that even if the optimization process takes more time than executing the original program once, the time investment can be worthy for widespread programs which may be executed in millions of devices, or systems for which critical components are executed millions of times. Furthermore, selecting the critical parts of the code to be optimized can aid to reduce the complexity of this approach.

\begin{algorithm}[!ht]
    \SetKwInOut{Input}{input}\SetKwInOut{Output}{output}\SetKw{lblendloop}{end\_loop:}\SetKw{goto}{goto}
    \Input{$\mu, \mathcal{S}_{\mathcal{P}}, M$}
    \Output{An optimized source code $\mathcal{S}_\mathcal{O}$}
    $\mathcal{S}_\mathcal{O} \leftarrow \mathcal{S}_{\mathcal{P}}$\;
    $\mathcal{O} \leftarrow C(\mathcal{S}_{\mathcal{P}})$\;
    $\tau_\mathcal{O} \leftarrow 0$\;
    \ForEach {$\alpha \in M$}
    {   
        $\tau_\mathcal{O} \leftarrow \tau_\mathcal{O}+t(\mathcal{O},\alpha)$\;
    }
    \ForEach{$m \in \mu$}
    {
        $\Omega \leftarrow mutate(m,\mathcal{S}_{\mathcal{P}})$\;
        \ForEach{$\mathcal{S}_{\mathcal{P}'} \in \Omega$}
        {
            $\mathcal{P}' \leftarrow C(\mathcal{S}_{\mathcal{P}'})$\;
            \If{$\mathcal{P}'== \varepsilon$\tcp{the program does not compile}}{\goto end\_loop}
            $\tau_{\mathcal{P}'}\leftarrow 0$\;
            $\epsilon \leftarrow$ {\bf true}\;
            \ForEach {$\alpha \in M$}
            {   
                $\epsilon \leftarrow \epsilon \; \& \;( out(\mathcal{O}, \alpha)==out(\mathcal{P}', \alpha))$\;
                \If{!$\epsilon$}{\goto end\_loop}
                $\tau_{\mathcal{P}'}\leftarrow \tau_{\mathcal{P}'}+t(\mathcal{P}',\alpha)$\;
            }
            \If{$\tau_\mathcal{O} > \tau_{\mathcal{P}'}$}
            {
                $\mathcal{S}_\mathcal{O} \leftarrow \mathcal{S}_{\mathcal{P}'}$\;
                $\mathcal{O} \leftarrow \mathcal{P}'$\;
                $\tau_\mathcal{O}\leftarrow \tau_{\mathcal{P}'}$\;
            }
            \lblendloop\;
        }
    }
    \caption{Code optimization using equivalent mutants}\label{algo}
\end{algorithm}

We propose Algorithm~\ref{algo} for source code optimization using equivalent mutants. Hereafter, $mutate$ denotes a mutation function which takes the mutation operator and the source code to mutate as parameters, and produces a set of mutants of the corresponding type. The resulting optimizations depend on the set of inputs $M$ on which the program is stimulated.   

Algorithm~\ref{algo} returns a source code which compiles to a program that is $M$-equivalent to the initial one. Therefore, for assuring the program equivalence, one can derive a set $M$ of inputs as a complete/exhaustive test suite which guarantees that the original and optimized programs have the same behavior \cite{mbt1}. In fact, the more \emph{precise} this set $M$ is constructed, the higher is the guarantee of the equivalence between the optimized and the original programs. 

\section{Preliminary Experimental Results}\label{experiments}
As a simple case study, we chose the source code of an intricate Java function which given a binary string, returns its integer value. Note that in this source code there is no verification that the string is indeed binary, however, we do not focus on such enhancements. The source code is shown below. 

\begin{lstlisting}[breaklines=true,basicstyle=\scriptsize,language=java]
static int b2tob10 (String binary) {
    String bin = new StringBuilder(binary).reverse().toString();
    int size = bin.length();

    if(bin.length() == 0)
        return 0;

    int pos = 1, i = 2, number = 0, count, aux;

    number += Integer.parseInt(bin.substring(0,1));

    while (i <= 1 << size - 1){
        aux = i;
        count = 0;
        while(aux > 0){
            count++;    
            aux = aux & (aux - 1);
        }       
        if (count > 1) {
            i +=2;      
            continue;   
        }       
            
        number += Integer.parseInt(bin.substring(pos, ++pos)) * i;
        
        i+= 2;  
    }   
    
    return number;
}
\end{lstlisting}

When performing experiments, Algorithm~\ref{algo} has been executed with the following parameters:
\begin{itemize}
    \item $\mu$Java as the mutation function,
    \item the source code $\mathcal{S_{\mathcal{P}}}$ as shown above, 
    \item $\mu=\{ROR,ASR,AOR\}$ as the mutation operator set,
    \item $M=\{111111111111111111111111110110,0,1\}$ as the set of inputs (test suite). 
\end{itemize}

The obtained set of optimized source code contains a mutant of interest, identified in $\mu$Java as \lstinline{ASRS_18} which is the replacement of an assignment operator (ASR), i.e., the statement \lstinline{i+=2} with \lstinline{i*=2}. The obtained mutant is $M$-equivalent to the original program as it outputs $1073741814, 0, 1$. At the same time, its overall execution time is $0.463s$, strictly less than the original program's overall execution time of $5.857s$. As it can be seen, the performance enhancement obtained by the showcased approach is significant. Furthermore, despite the fact that $M$ is not a complete test suite, one can assure that the obtained mutant is equivalent to the initial program. Indeed, the variable \lstinline[language=java]{number} (the return value of the function) gets updated only when the \lstinline[language=java]{continue} instruction is not executed. The \lstinline[language=java]{continue} instruction is not executed under the condition that there exists more than one `1' in the binary representation of the iterator \lstinline[language=java]{i}. Any binary string with only one `1' represents a power of two; that implies that the next time the condition does not execute the \lstinline[language=java]{continue} instruction occurs when the iterator \lstinline[language=java]{i} equals \lstinline[language=java]{i*2}.

As the Java compiler~(\lstinline{javac}) and virtual machine~(\lstinline{JVM}) perform static and dynamic optimizations, the previously presented optimization outperforms the optimizations performed by both, \lstinline{javac} and \lstinline{JVM}. However, in order to compare this approach to \emph{traditional} compiler optimizations, we translated the example to standard C code. The overall running time of the program obtained from compiling the original C source code without optimizations was $14.596s$. The program obtained by compiling the program with the highest Gnu Compiler Collection~(gcc) optimizations (\lstinline{gcc -O3}) had an overall running time of $4.878s$. The program obtained by compiling the mutant without any compiler optimizations had an overall running time of $0.022s$. As it can be seen, the provided optimizations outperform the traditional ones. The main reason behind this improvement is that the optimizations obtained using equivalent mutants affect the semantics of the source code, differently from compiler optimizations.

\section{Conclusion}\label{conc}
We presented an approach for source code optimization using equivalent mutants. Preliminary experimental results show that the presented approach can outperform the traditional compiler optimizations, whenever the approach is applicable. Many directions are left open for future work and perhaps the most important of them is the study of the applicability of the approach, by performing a thorough experimental evaluation. Other interesting directions include studying other types of source code optimization together with the extended list of mutation operators and the exploration of symbolic model checking for the efficient verification of equivalent mutants. 




\section*{References}
\bibliographystyle{elsarticle-num} 
\bibliography{refs}


\end{document}